\documentclass{PoS}

\usepackage[mathscr]{eucal} 
\usepackage{amsmath}
\usepackage{amsthm} 
\usepackage{xcolor}  

\usepackage{sidecap} 
\usepackage{wrapfig}    
\usepackage{subfig} 
\usepackage{float}

\title{4d ensembles of percolating center vortices and chains} 

\ShortTitle{4d ensembles of percolating center vortices and chains}

\author{\speaker{Luis E. Oxman} \\
        Fluminense Federal University\\
        E-mail: \email{leoxman@id.uff.br}}  

\abstract{ In this work, we review a recently proposed  measure to compute center-element averages in a mixed ensemble of center vortices and chains with non-Abelian d.o.f. and monopole fusion. When center vortices percolate and monopoles condense,  the average is captured by a saddle point and collective modes in a YMH model. In this manner, the L\"uscher term, confining 
flux tubes with N-ality and confined gluons were accommodated in an ensemble picture.}

\FullConference{XIII Quark Confinement and the Hadron Spectrum - Confinement2018\\
		31 July - 6 August 2018\\ 
		Maynooth University, Ireland}

\begin{document}

\section{Introduction}

\begin{figure} 
\centering    
\subfloat[Small defects]{ \includegraphics[scale=.16, bb= -2cm -7cm 20cm 3cm]{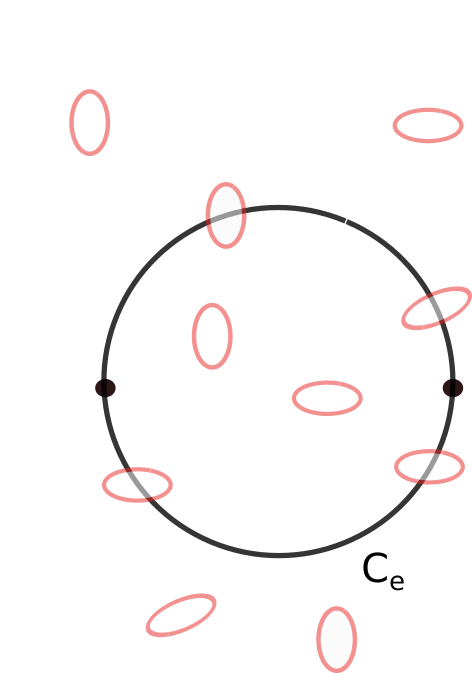}}  
  \hspace{1.8cm}  
	 \subfloat[Percolating defects]{\includegraphics[width=.19\textwidth]{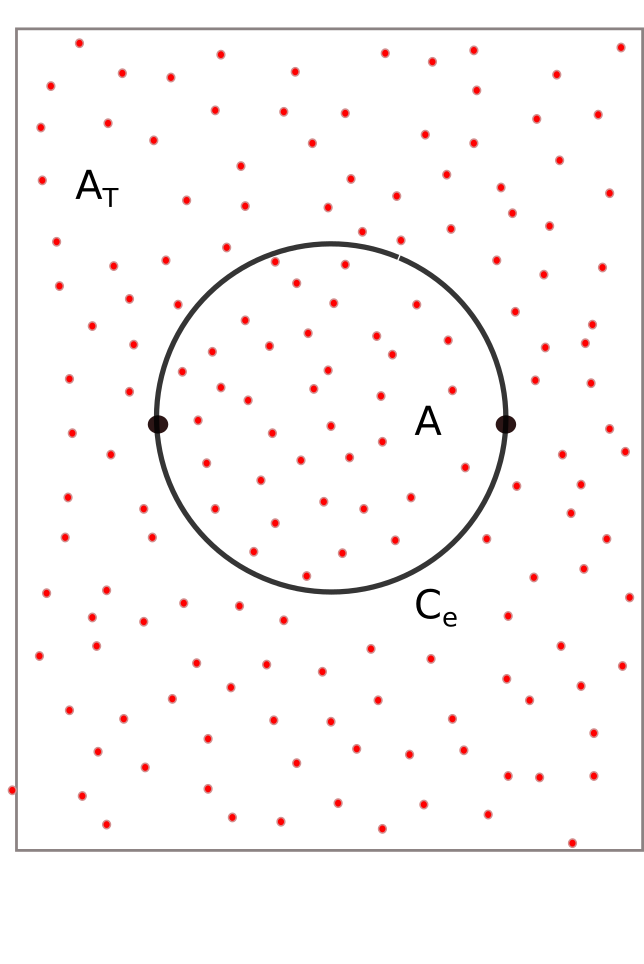}} \hspace{.2cm}  
\caption{\small Center-vortex worldlines (in red) and Wilson loop ${\mathcal C}_{\rm e}$ (in black) in a 3d Euclidean spacetime.  }  
\label{conj1}  
\end{figure}

\begin{figure}     
\centering 
\subfloat[]{ \includegraphics[scale=.2]{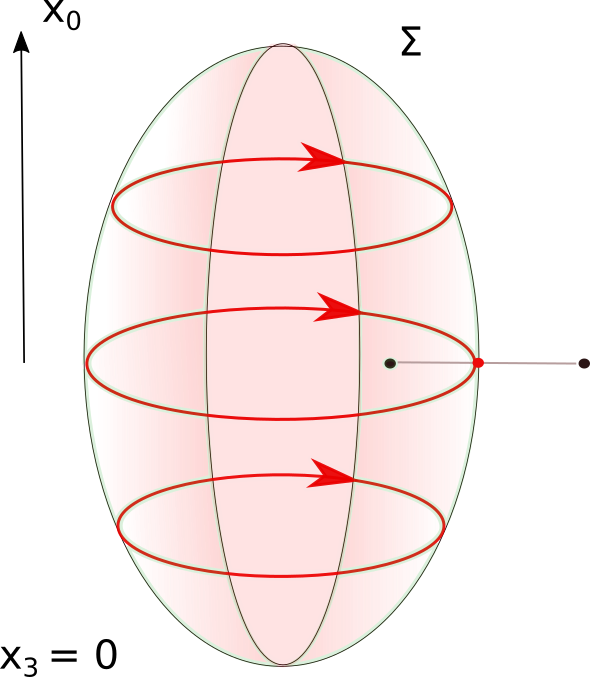}} \hspace{1.8cm}   
	 \subfloat[]{\includegraphics[scale=.22]{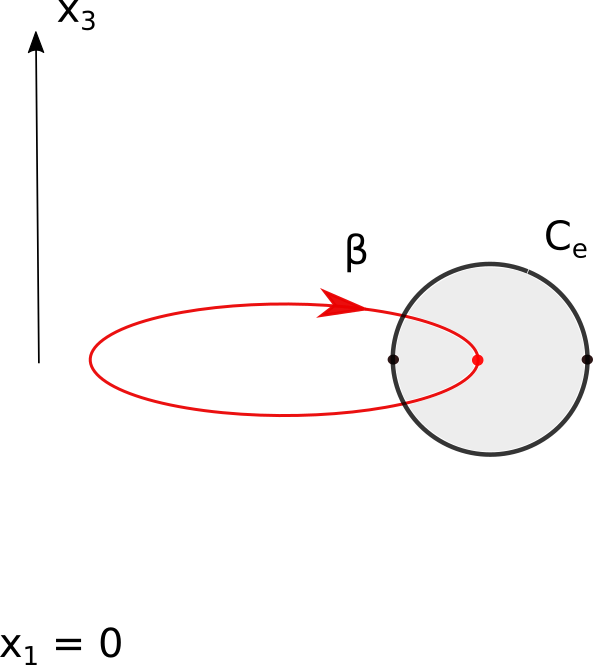}} \hspace{.2cm}  
\caption{\small Center-vortex worldsurface $\Sigma$ in the hyperplane $x_3=0$. It represents the creation propagation and annihilation of a one-dimensional oriented object. The pair of points (in black) and the segment between them  belong to $\cal{C}_{\rm e}$ and  $S(\cal{C}_{\rm e})$, respectively (a). The complete  $\cal{C}_{\rm e}$ can be seen on the hyperplane $x_1=0$ (b).  }    
\label{CV-link}     
\end{figure}

Monte Carlo simulations provide a direct computational path that goes from the large quantum fluctuations in $SU(N)$ Yang-Mills theories to observables. The underlying physical mechanism for  confinement should accommodate the large amount of information obtained this way. In a  $d$-dimensional Euclidean spacetime, the observed asymptotic potential between fundamental quarks \cite{Bali} contains a  linear term plus a universal  
L\"uscher correction  $-\frac{\pi (d-2)}{24 R}$ \cite{LW} that point to the formation of a confining string. Moreover, the field distribution around the string reveals a flux tube structure (see  \cite{Cosmai-2017} and refs. therein). Another important property  
is that the potential $ V_{\rm D}(R)$ between quarks in an irreducible $\mathscr{D}$-dimensional representation ${\rm D}$ displays $N$-ality. That is, the asymptotic  string tension extracted from the Wilson loop
\begin{gather} 
{\mathcal W}_{\rm e}[\mathscr{A}]  =  \frac{1}{\mathscr{D}} \, {\rm tr}\, {\rm D}  \left( P \left\{ e^{i \oint_{{\cal C}_{\rm e}} dx_\mu\,  A_{\mu}(x)  } \right\} \right)  \makebox[.5in]{,} \langle {\mathcal W}_{\rm e}[\mathscr{A}] \rangle \sim e^{-T\,  V_{\rm D}(R)} \;, 
\end{gather}  
 depends on how the center of $SU(N)$ is realized \cite{dFK}, 
\begin{equation}
e^{i \frac{2\pi}{N}} I \to {\mathrm D} \left(e^{i \frac{2\pi}{N}} I \right)  \;.
\label{celem}
\end{equation}
This favors a  mechanism based on percolating center vortices \cite{tHooft:1977nqb}-\cite{reinhardt-engelhardt}. In 3d (resp. 4d) Euclidean spacetime, the field strength for thin center vortices  is localized on closed worldlines (resp. worldsurfaces) while the gauge field is locally, but not globally, a pure gauge. This
yields
\begin{equation}
  {\mathcal W}_{\rm e}[\mathscr{A}_{\rm center-vortex}] = \left[  e^{- i2\pi\,  \vec{\beta}  \cdot   \vec{w}_{\rm e} } \right]^{L({\rm c.v.}, \cal{C}_{\rm e})} \makebox[.5in]{,} \vec{\beta} = 2N\, \vec{w}\;,
  \label{proper}   
\end{equation}  
where $L({\rm c-v.}, \cal{C}_{\rm e})$ is the total linking number between the defects and the Wilson loop $\cal{C}_{\rm e}$, which can be equated to the intersection number with a surface $S(\cal{C}_{\rm e})$ whose border is  $\cal{C}_{\rm e}$. The $(N-1)$-tuples $\vec{w}$ and $\vec{w}_{\rm e}$ are weights of the fundamental representation and of the quark representation ${\rm D}$, respectively.
 If the ensemble were formed by small defects, the average $\langle {\mathcal W}_{\rm e}[\mathscr{A}] \rangle $ would lead to a perimeter law. In fact, the center vortices detected in the lattice form an ensemble of 
large percolating defects with a density that scales towards a physical finite value in the continuum
\cite{Langfeld:1997jx, DelDebbio:1998luz}.  In this case, the occurrences with nontrivial linking scale as the area of the Wilson loop. In the three-dimensional case, this picture is shown in Fig. \ref{conj1}.   
In four dimensions, the situation is similar, with the difference that the field strength is localized on (large percolating) closed worldsurfaces, which can link the Wilson loop (see Fig. \ref{CV-link}). 
This, together with Eq. \eqref{proper}, implies an area law with $N$-ality. 
However, neither L\"uscher terms nor flux tubes have been observed in center-vortex ensembles.  In addition,  although the implied asymptotic adjoint string tension vanishes, it would be important 
to reproduce the complete adjoint string-breaking process. That is, at some  distance the energy stored in the adjoint string should give place to screening, due to the formation of gluonic excitations around the external sources. 

\begin{figure} 
\centering       
	 \subfloat{\includegraphics[scale=.4]{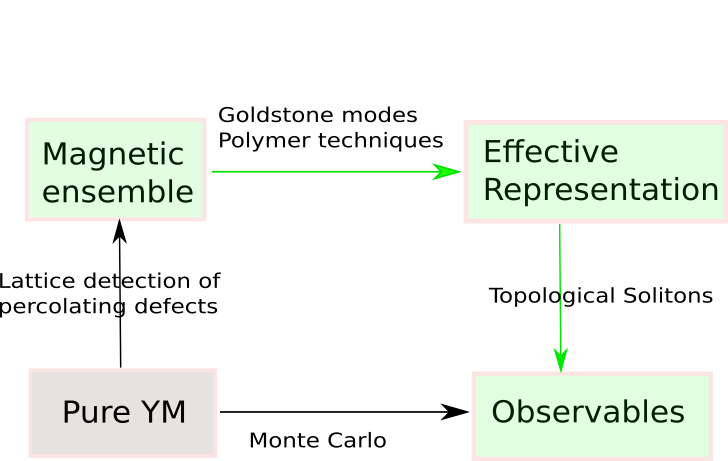}}  
\caption{}
\label{diag} 
\end{figure}

Besides center vortices, chains are among the candidates that could capture the infrared quantum fluctuations in the YM vacuum. They are formed by center-vortex branches attached in pairs to monopoles. Indeed,  in lattice simulations, they account for 97\% of the cases \cite{Ambjorn:1999ym}. In a recent work  
\cite{mixed}, we studied 4d mixed ensembles composed by center vortices and chains. In particular, we motivated the inclusion of non-Abelian magnetic degrees of freedom (d.o.f.) as well as monopole fusion, and showed how to deal with an ensemble of two-dimensional defects.  
All these ingredients were essential to obtain an effective description that can capture  the whole physical picture of confinement. When computing center element averages, they lead to a saddle point based on topological solitons in an effective $SU(N) \to Z(N)$ SSB model, corrected by collective coordinates. In what follows, we shall review these ideas by following the 
path (in green) summarized in Fig. \ref{diag}.

\section{A measure for percolating center vortices in the lattice}

\subsection{Three dimensions}  
 
In a planar system, center vortices are localized on points, so they are created or annihilated by a field operator $\hat{V}(x)$. The emergence of this order parameter can be clearly seen by applying polymer  techniques  to center-vortex worldlines  \cite{deLemos:2011ww}. In Ref. \cite{Oxman-Reinhardt-2017}, we showed
that the 3d ensemble average of center elements (cf. Eq. \eqref{proper}) over closed worldlines with tension $\mu_{\rm v}$,  small positive stiffness $1/\kappa_{\rm v}$, and excluded volume effects is 
\begin{gather}
\big\langle \big(e^{i \frac{2\pi}{N}}\big)^{L({\rm c-v.}, \cal{C}_{\rm e})} \big\rangle = \frac{Z_{\rm v}^{(3)} [s_\mu]}{Z_{\rm v}^{(3)} [0]} \makebox[.3in]{,} Z_{\rm v}^{(3)} [s_\mu]
=\int [{\cal D}V][{\cal D}\bar{V}]\,   e^{-\int
	d^{3}x\,\left[ \frac{1}{3 \kappa}\,  \overline{D_\mu V} D_\mu V +\frac{1}{2\zeta}\, (\overline{V} V -v^2)^2  \right]} \label{avecent}  \;, \\
D_\mu = \partial_{\mu}- i \frac{2\pi }{N} s_\mu \makebox[.5in]{,} v^2 = -\mu_{\rm v} \zeta \makebox[.5in]{,} \zeta > 0\;. 
\end{gather} 
The gauge field $s_\mu$, localized on $S(\cal{C}_{\rm e})$, is required to write linking numbers as
intersection numbers in the initial ensemble.
 For small worldlines ($\mu_{\rm v} > 0$) we must deal with the complete complex field $V$. On the other hand, for large percolating worldlines ($\mu_{\rm v} < 0$),  the soft degrees of freedom are the Goldstone modes $\gamma$, $V(x) \approx v \, e^{i\gamma (x)}$. In this case, we switched to the lattice 
and discussed the Goldstone fluctuations by relying on the frustrated 3d XY model
\begin{eqnarray}  
Z_{\rm v}^{{\rm latt} } [\alpha_\mu]
\approx \int [{\cal D}\gamma] \,   e^{-S_{(3)}^{\rm latt}} \makebox[.5in]{,} S_{(3)}^{\rm latt} = \tilde{\beta} \, \sum_{x, \mu }  \mathrm{Re} \left[ 1- 
 e^{i \gamma(x + \hat{\mu})} e^{-i \gamma(x)} e^{-i \alpha_\mu ({x})}   \right] \;,
\label{xym}
\end{eqnarray} 
where $e^{ i \alpha_\mu (\mathbf{x})} = e^{i \frac{2\pi}{N}}$, if  $S
 ({\cal C}_{\rm e})$ is crossed by the link, and it is $1$ otherwise. As is well-known, there is a critical point at $\tilde{\beta}_{\rm c}  \approx 0.454 $ where the continuum limit with large percolating worldlines is recovered. At that point, a Wilson loop area law with $N$-ality was obtained.
 
\subsection{Four dimensions}  
     
In the $3+1$ dimensional world, as center vortices are one-dimensional objects spanning closed worldsurfaces,
the emergent order parameter is a string field. Nonetheless, by focusing on percolating worldsurfaces, the desired generalization can be drastically simplified. Let us recall how the 3d center-vortex ensemble is  codified in the final lattice representation   
\eqref{xym}. Expanding in powers of $\tilde{\beta}$,  the contribution is originated from sets of links $\{ (x, \mu) \}$ that form closed one-dimensional arrays.  Otherwise, the integral over the configuration 
$\gamma$-space  would vanish. Furthermore, the arrays are accompanied by a center element whenever they intersect $S(\cal{C}_{\rm e})$, so they can be identified with the center-vortex worldlines. Back to four dimensions,  the string field for an Abelian condensate of closed strings is known to be $V({\rm string}) \approx v \, e^{i\gamma ({\rm string})}$, with the possible phases \cite{Rey}
\[ 
\gamma_\Lambda({\rm string}) = \oint_{\rm string} dx_\mu \, \Lambda_\mu   \;.
\]
In other words, the Goldstone modes in this case are gauge fields. In the condensate,  the lattice string field theory was  approximated by a field theory: the $U(1)$ gauge-invariant Wilson action. 
In Ref. \cite{mixed},  motivated by these considerations, we proposed the ensemble measure   
\begin{gather}
 Z_{\rm v}^{\rm latt} [\alpha_{\mu \nu}] = \int [{\cal D}V_\mu]\,  e^{- S_{(4)}^{\rm latt}}  \makebox[.3in]{,} 
S_{(4)}^{\rm latt} = \tilde{\beta} \,\sum_{\mathbf{x}, \mu < \nu } \mathrm{Re}\;  {\rm tr} \left[ I -  V_\mu(x) V_\nu(x + \hat{\mu}) V^\dagger_\mu(x + \hat{\nu}) V^\dagger_\nu(x) \,   e^{-i\,  \alpha_{\mu \nu} }  \right] 
\label{cv-measure} 
\end{gather}   
and associated $Z_{\rm v}^{\rm latt} [\alpha_{\mu \nu}]/Z_{\rm v}^{\rm latt} [0]$ with a center element average in the lattice.   
The frustration  $\alpha_{\mu \nu}$ is nonzero on plaquettes $x,\mu,\nu$ that intersect $S(\mathcal{C}_{\rm e})$, where it gives 
$ e^{-i\,  \alpha_{\mu \nu} }  =  e^{- i2\pi\,  \vec{\beta}  \cdot   \vec{w}_{\rm e} }$.  Besides the Abelian measure based on $V_\mu \in U(1)$, there is another natural one based on $V_\mu \in SU(N)$ (see section \ref{nonab}). In both cases, when expanding in powers of $\tilde{\beta}$,  the relevant terms correspond to plaquette configurations distributed on closed surfaces, which can form singlets on every link. When the surface is linked by $\cal{C}_{\rm e}$, a center-element ($  e^{\pm i  \alpha_{\mu \nu} } $) for quarks in representation ${\rm D} $ is generated.  This measure is in complete analogy with the 3d case.
In Fig. \ref{CV-link}, we show the intersection point (in red) between a linked center-vortex
worldsurface and $S(\mathcal{C}_{\rm e})$. This is done on the hyperplanes at $x_3=0$ (Fig. \ref{CV-link}a) and at $x_1=0$ (Fig. \ref{CV-link}b), where we only see a segment belonging to $S(\mathcal{C}_{\rm e})$
and a center-vortex string, respectively.

\section{Center vortices and chains with non-Abelian d.o.f.}  
\label{nonab}  

In the lattice, by considering the lowest eigenfunctions of the adjoint covariant Laplacian, a procedure to detect magnetic defects was proposed in Refs. \cite{faber, deF-P}.  In Ref. \cite{OS-det},  relying on  auxilliary adjoint fields that satisfy a classical equation of motion, we implemented this idea in the continuum. In particular, we defined a mapping from pure YM configurations $\mathscr{A}_\mu$ into  $S[\mathscr{A}] \in SU(N)$ with the property,  
\[ 
S  [\mathscr{A}^{U_{\rm e} }] = U_{\rm e}  S[\mathscr{A}]   \makebox[1.9in]{(regular $U_{\rm e}$).}
\]
where $\mathscr{A}^{U_{\rm e}}$ denotes a chromoelectric gauge transformation.  
Although $\mathscr{A}_\mu$ is regular,  $S[\mathscr{A}] $ will generally contain defects. 
This induces a partition into classes:  $\mathscr{A}_\mu$ and  $\mathscr{A}'_\mu$ are in the same sector iff there is a regular $U_{\rm e}$ such that $ S[\mathscr{A}']= U_{\rm e}\, S[\mathscr{A}]$.    
Configurations in different sectors cannot be physically equivalent. On each sector, there are infinitelly many physically inequivalent fields which can be gauge fixed by choosing variables such that $A_\mu \to  {\rm Ad}(S_0) $, where $S_0$ is a class representative. 
In the continuum we can clearly perceive the presence of non-Abelian degrees of freedom \cite{mixed}. If 
$S[\mathscr{A}]$ contains defects, gauge fields $\mathscr{A}'_\mu$ such that $S[\mathscr{A'}] = S[\mathscr{A}]\,  \tilde{U}^{-1}$ (regular $\tilde{U}$) have defects at the same location but generally belong to different sectors.
\begin{figure}
\centering 
\subfloat[]{ \includegraphics[scale=.2]{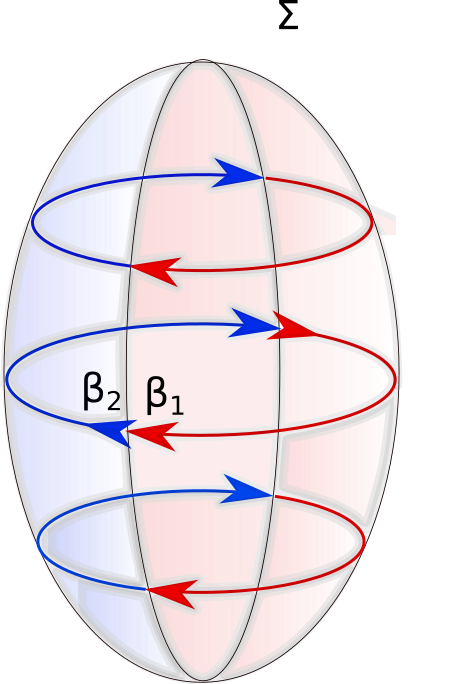}} \hspace{.7cm}   
	 \subfloat[]{\includegraphics[scale=.2]{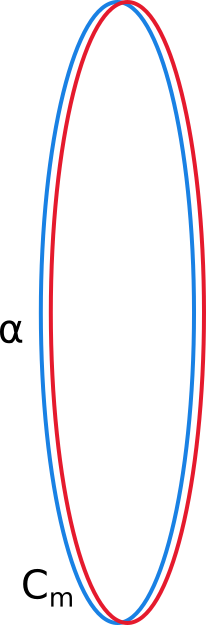}} \hspace{2cm}    
	 \subfloat[]{ \includegraphics[scale=.2]{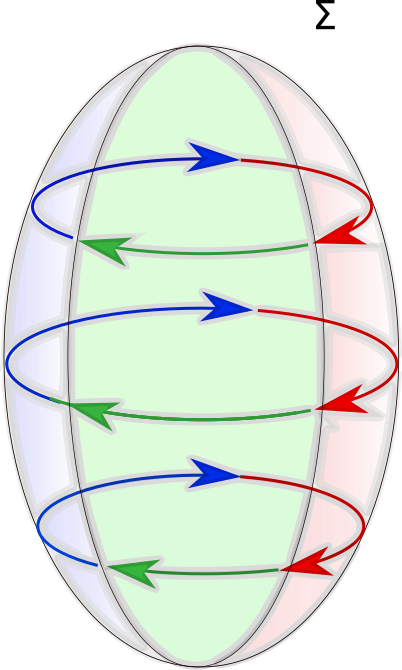}} \hspace{.7cm}    
	 \subfloat[]{\includegraphics[scale=.2]{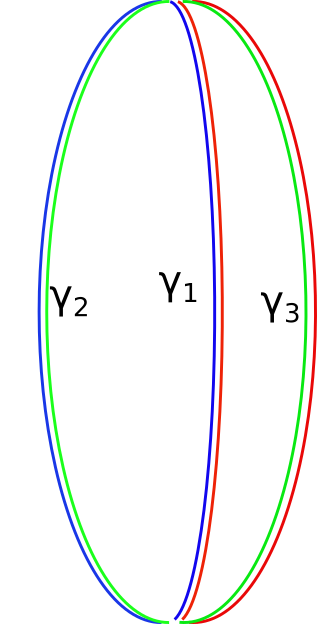}} 
\caption{\small a) worldsurface spanned by a stringlike chain with branches carrying fundamental weights  
$\vec{w}_1$ (in blue), $\vec{w}_2$ (in red). The associated monopole worldline with adjoint weight $\vec{w}_1 -\vec{w}_2$ is shown in b). In c) there is a third branch carrying fundamental weight $\vec{w}_3$ (in green). d) three-line monopole fusion. }     
\label{with-m}     
\end{figure}  
  
Among the magnetic defects, besides the oriented center vortices, we payed special attention to chains. This is 
because the detection of center vortices in 4d lattice Yang-Mills theory showed that 
in most cases they contain monopoles \cite{Ambjorn:1999ym}. In the continuum, the Lie algebra flux orientation changes at the monopoles \cite{Reinhardt:2001kf}.  Thus, the closed monopole worldline 
$\mathcal{C}_{\rm m}$ carries a weight of the adjoint representation (the difference of fundamental weights).
This is illustrated in Figs. \ref{with-m}a and \ref{with-m}b. Similarly to  $\mathscr{A}_{\rm center-vortex}$, thin chain configurations $\mathscr{A}_{\rm chain}$ can also be locally (but not globally) written as a pure gauge. The associated $SU(N)$-mapping must contain a sector outside the Cartan subgroup that changes from 
the identity map to a Weyl transformation when we go from one branch to the other. In Fig. \ref{with-m}c we show the creation, propagation,  and annihilation of a string-like chain with three different fundamental weights ($N \geq 3$).  The associated monopole creation-annhilation process is depicted in Fig. \ref{with-m}d.  
This and other rules involving four-lines, as well as three-line fusion 
in the $\mathfrak{su}(2)$-subalgebras ($N \geq 2$), were discussed in Ref. \cite{mixed}.

 \section{Chains and monopole fusion in the lattice}

\begin{figure} 
\centering  
\subfloat[]{ \includegraphics[scale=.15]{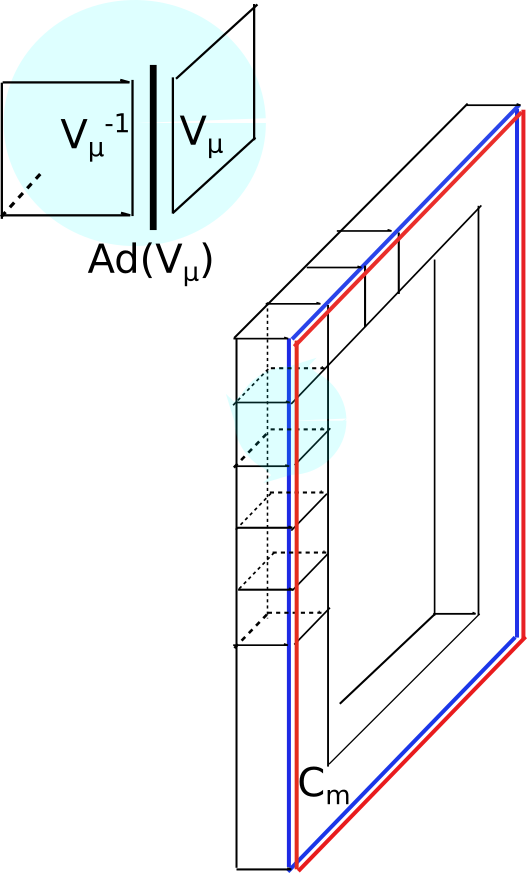}} \hspace{1.5cm}    
	 \subfloat[]{\includegraphics[scale=.15]{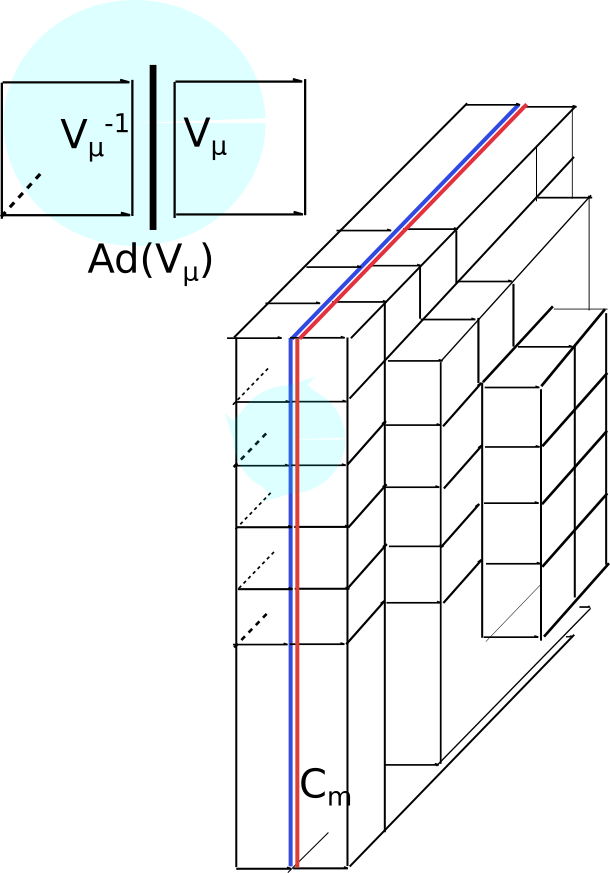}} \hspace{1.3cm}  
	  \subfloat[]{\includegraphics[scale=.15, bb= -2cm 3cm 20cm 3cm]{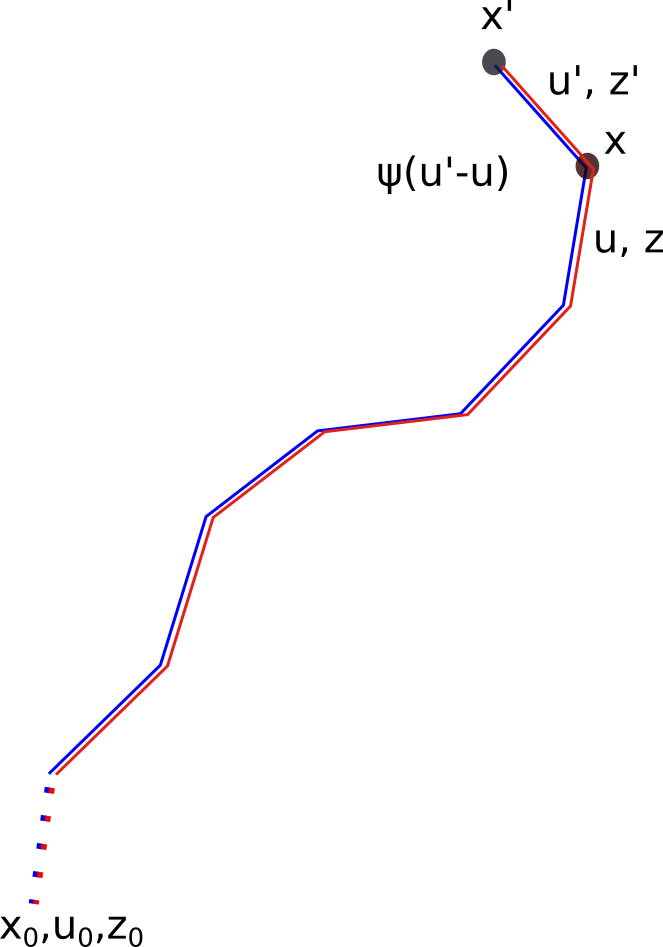}}   
\caption{\small Chains  in the lattice with an adjoint Wilson loop ${\cal C}_{\rm m}$ (red-blue lines). We show a configuration that contributes to the lowest order in $\tilde{\beta}$ (a), and one that becomes more important as $\tilde{\beta}$ is increased (b).   }      
\label{mon-curr}    
\end{figure}

For thin defects, neither non-Abelian magnetic d.o.f. nor monopole worldlines affect the Wilson loop:  ${\mathcal W}_{\rm e}[\mathscr{A}_{\rm thin}]$ continues to be the center element on the right-hand side of Eq. \ref{proper}, with the linking number computed between the thin defect and $\mathcal{C}_{\rm e}$. 
On the other hand, the YM average of ${\mathcal W}_{\rm e}$ amounts to a sum over $S_0$ weighted 
by the path-integral over the gauge fields in each sector, with appropriate conditions so as to work with a regular $A_\mu$. The latter depend on the different degrees propagated on top of center vortices, so we suggested that the YM ensemble measure could have a nontrivial dependence on them. The non-Abelian magnetic d.o.f. led us to consider a center-vortex measure \eqref{cv-measure} with $V_\mu \in SU(N)$. In turn, this led us to propose a 4d ensemble measure for chains. Center-vortex branches were naturally attached to closed monopoles worldlines by means of dual adjoint Wilson loops. That is,  we included in the ensemble terms of the form 
\begin{gather}
Z^{\rm latt}_{\rm mix} [s_{\mu \nu}]\big|_{\rm p} \propto  \int [{\cal D} V_\mu] \, e^{-\tilde{\beta}  \sum_{\mathbf{x}, \mu < \nu } \mathrm{Re}\;  {\rm tr} [ I -  V_\mu(x) V_\nu(x + \hat{\mu}) V^\dagger_\mu(x + \hat{\nu}) V^\dagger_\nu(x) e^{-i \alpha_{\mu \nu}(x)  }  ]  } \,  {\mathcal W}^{(1)}_{\rm Ad} \dots  {\mathcal W}^{(n)}_{\rm Ad} \;,
\nonumber \\ 
{\mathcal W}^{(k)}_{\rm Ad}    =  \frac{1}{N^2-1} \, {\rm tr}\,   \Big( \prod_{(x,\mu)\in \,
{\cal C}^{\rm latt}_k } {\rm Ad} \big( V_{\mu}(x) \big)   \Big)  \;.   
\label{w-latt} 
\end{gather}
On the one hand, the Wilson loops implement the adjoint charge and non-Abelian d.o.f. carried by monopoles. On the other, expanding in powers of $\tilde{\beta}$, in order to produce singlets, the plaquettes must be attached in pairs to the links of ${\cal C}^{\rm latt}_k $. Otherwise, the integral of an isolated factor ${\rm Ad }(V_\mu)|_{AB}$ would vanish 
(see Fig. \ref{mon-curr}). 
Closed arrays with monopole fusion were similarly generated using holonomies along $n$ fused lines that form  gauge invariant objects. For example, to produce arrays like the one in Fig. \ref{with-m}c, we considered
holonomies along $\gamma_1$, $\gamma_2$, $\gamma_3$ (see Fig. \ref{with-m}d), and included partial contributions such as
\begin{gather}
 Z^{\rm latt}_{\rm mix} [s_{\mu \nu}]\big|_{\rm p} \propto  \int [{\cal D} V_\mu] \, e^{-\beta  \sum_{\mathbf{x}, \mu < \nu } \mathrm{Re}\;  {\rm tr} [ I -  V_\mu(x) V_\nu(x + \hat{\mu}) V^\dagger_\mu(x + \hat{\nu}) V^\dagger_\nu(x) e^{-i \alpha_{\mu \nu}(x)  }  ]  } \, D_3
\nonumber \\
D_3(\gamma_1, \gamma_2, \gamma_3)  =  F_{A_1 A_2 A_3} \, F_{B_1 B_2 B_3} \, {\rm Ad}(\Gamma_1)|_{A_1 B_1}\, {\rm Ad}(\Gamma_2)|_{A_2 B_2} {\rm Ad}(\Gamma_3)|_{A_3 B_3} \;,
\label{tres}
\end{gather}  
where $F_{ABC}$ is an (adjoint) group invariant tensor.  Other partial contributions can be generated with products of $D_3$-blocks computed along different line-triplets. Based on the ingredients above, there are many possibilities to define a 4d lattice measure. They depend on the fusion rules, the choice of invariant tensors, etc.

\section{Effective fields in the continuum}

In Ref. \cite{mixed}, we weighted the lattice contributions with phenomenological properties, and assumed the presence of a critical point where the continuum limit is attained. Taking  $ V_\mu(x) = e^{ia \Lambda_\mu(x)} $,  $ \Lambda_\mu \in \mathfrak{su}(N) $,  with a dual coupling ($  \frac{1}{g^2} \sim \tilde{\beta} $), and  using  $e^{- i2\pi\,  \vec{\beta}  \cdot   \vec{w}_{\rm e} }\, I  
 =  e^{-i\, 2\pi\,  \vec{\beta}_{\rm e}|_q T_q}$, the naif continuum limit of the proposed center-element average  
is\footnote{$T_q$, $q=1, \dots, N-1 $, are Cartan generators, and $\vec{\beta}_{\rm e} = 2N\, \vec{w}_{\rm e}$.} 
\begin{gather}
 \frac{Z_{\rm mix} [s_{\mu \nu}]}{Z_{\rm mix} [0]} 
\makebox[.5in]{,}   Z_{\rm mix} [s_{\mu \nu}] = \int [{\cal D}\Lambda_\mu] \,   e^{-\int d^4x\,    \frac{1}{4g^2} \,   \left( F_{\mu \nu}(\Lambda ) - 2\pi s_{\mu \nu} \vec{\beta}_{\rm e} \cdot \vec{T}  \right)^2  }
Z_{\rm mon} [\Lambda] \;, 
\label{mixe}
\end{gather} 
where $s_{\mu \nu}$ is localized on $S(\cal{C}_{\rm e})$. $Z_{\rm mon} [\Lambda] $ collects the contribution of dual Wilson loops and holonomies ${\rm Ad} \left( \Gamma[\Lambda] \right)$  weighted with tension $\mu$ and stiffness $1/\kappa$.  The analysis of a monopole worldline that grows in the presence of a non-Abelian gauge field showed that $ Q(x,u,x_0,u_0,L)$, the weighted and integrated holonomy over worldline shapes, 
satisfies a Fokker-Planck diffusion equation in $(x,u)$-space 
(see also Ref. \cite{GBO}). The integrated paths have fixed length $L$, initial and final positions $x_0 , x \in \mathbb{R}^4$, and tangent vectors 
$u_0,u \in S^3$ (see Fig. \ref{mon-curr}c).  For small stiffness and at large distances, we obtained  
\begin{gather}  
 \partial_L Q(x,x_0,L) \approx -O\, Q(x,x_0,L)   \makebox[.3in]{,}  O = - \frac{\pi}{12 \kappa} \, \left( \partial_{\mu} - i\, {\rm Ad} \big(\Lambda_{\mu} \big) \right)^2  + \mu \;.  
 \end{gather} 
In particular, counting both loop orientations, and integrating all possible loop lengths, a diluted monopole gas provides a factor 
\begin{gather}     
Z_{\rm gas} [\Lambda] 
\approx e^{\, - {\rm Tr} \,  \ln O  }    
 = ( {\rm Det} \, O )^{-1}  \;.
 \label{gasm2}  
\end{gather}
For  $N_{\rm F}$ real adjoint flavors,  the factor is $ \big(Z_{\rm gas} [\Lambda]\big)^{N_{\rm F}/2} $.  Scenarios with $N_{\rm F}=N(N-1)$  (resp. $N_{\rm F} = N^2-1$) that include fusion rules in the Cartan subalgebra (resp. Cartan and $\mathfrak{su}(2)$ subalgebras) are among the possibilities. Additionally, in any array of fused lines each path-integral over line-shapes $Q(x,x_0,L)$ has to be integrated over $L$. This gives a Green's function $G(x,x_0)$, $O\, G(x,x_0) = \delta(x-x_0  ) \, I_{\mathscr{D}_{\rm Ad}} $ for every holonomy. For example, the contribution originated from Eq. \eqref{tres} is
\begin{gather} 
 C_3 \propto  \int d^4x \, d^4x_0\,  F_{A_1 A_2 A_3} \, F_{B_1 B_2 B_3} G(x,x_0) |_{A_1 B_1 }  
 G(x,x_0) |_{A_2 B_2 }  G(x,x_0) |_{A_3 B_3 } \;.
\end{gather}
Thus, including all possible higher-order combinations, the average of center elements
was represented by Eq. \eqref{mixe} together with 
\begin{gather}  
 Z_{\rm mon} [s_{\mu \nu}] =  \int [{\cal D }\psi ] \,   e^{-\int d^4x\,  \left[    \frac{1}{2} (D_\mu \psi_I , 
 D_\mu \psi_I ) +  V_{\rm H}(\psi)  \right]  } \makebox[.5in]{,}  D_\mu(\Lambda)\, \psi  = \partial_{\mu} \psi - i\, [\Lambda_{\mu} ,  \psi ] 
 \end{gather}  
($I=1, \dots, N_{\rm F}$). For example, when $F_{ABC}$ are the antisymmetric $\mathfrak{su}(N) $-structure constants, we have
\[
V_{\rm H}(\psi) =  m^2   (\psi_I, \psi_I ) + \gamma (\psi_{1} , \psi_{2} \wedge \psi_3] ) + \dots  
\makebox[.5in]{,}  m^2 = (12/\pi) \,  \mu \kappa  \;,
\] 
where the dots represent other cubic and quartic interactions ($X \wedge Y = -i [X,Y]$). The couplings weight the abundance of each fusion type. Positive stiffness and negative tension  implies $m^2 < 0$, so that the effective model can easily display S$U(N) \to Z(N)$ SSB\footnote{In fact, because of the cubic terms, this pattern can also occur for small positive $m^2$ values.}, that is, a vacuum manifold ${\cal M}=  {\rm Ad}(SU(N))$ with  $\Pi_1 ({\cal M}) = Z(N)$. In the lattice, this pattern would be attained at a critical point in the Fradkin-Shenker phase-diagram for gauge theories coupled to adjoint flavors \cite{Eduardo}. From the ensemble point of view, this is expected to correspond to percolating center vortices, as they are described by Goldstone modes in a condensate (the dual gauge fields), as well as percolating monopoles, as 
positive stiffness and negative tension favors the proliferation of large worldlines.

\section{Discussion} 

We reviewed a recently proposed path that leads from 4d Yang-Mills magnetic ensembles to effective field confining models. In Ref. \cite{mixed}, we introduced a (lattice) measure to compute center-element averages over percolating worldsurfaces spanned by center vortices and center-vortex branches attached to monopoles (chains). These configurations have been observed in lattice simulations of $SU(N)$ pure YM theory. In that reference, we also motivated the presence of non-Abelian magnetic degrees of freedom and monopole fusion
in these ensembles. In the lattice, the center-vortex worldsurfaces were then generated by a dual YM action, while the monopole worldlines were generated by dual holonomies. The monopole sector was integrated using polymer techniques. This led to emergent adjoint fields and effective cubic and quartic Feynman vertices. When the monopole worldlines also percolate, we made contact with field models displaying S$U(N) \to Z(N)$ SSB, which have  been extensively studied in the literature (see Refs. \cite{Fidel2}-\cite{conf-qg} and references therein). At asymptotic distances, the center-element average can then be computed by a saddle-point with $N$-ality. For nontrivial $N$-ality, this corresponds to a stringlike object corrected by transverse quantum fluctuations, which originate the L\" uscher term.  
\begin{figure} 
\centering     
	  	 \subfloat[]{ \includegraphics[scale=.1, bb=2.3cm 0 0 0]{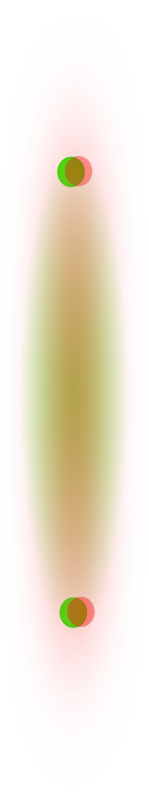} }  \hspace{3.7cm}    	     
	  	 \subfloat[]{ \includegraphics[scale=.11]{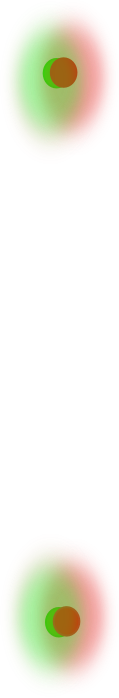} }  	
\caption{ }      
\label{adjunta}    
\end{figure} 
For adjoint quarks, at intermediate distances there is also a confining string (Fig. \ref{adjunta}a). 
However, at asymptotic distances the energy is minimized by the creation of a dual monopole around each source (valence gluon), see Fig. \ref{adjunta}b. Moreover, valence gluons are confined, as $\Pi_2({\cal M}) = 0$ precludes the presence of isolated topological pointlike solitons. \vspace{.2cm}
  
Summarizing, 4d mixed ensembles of center vortices and chains can accommodate the complete observed Monte Carlo phenomenology, when they carry non-Abelian magnetic d.o.f. and have appropriate fusion rules. 
The detection of these properties in $SU(N)$ pure YM lattice simulations will be important to support this scenario. By now, theoretical investigations aimed at determining the optimal effective field description and the corresponding ensemble are underway.

\section*{Acknowledgements}

 The Conselho Nacional de Desenvolvimento Cient\'{\i}fico e Tecnol\'{o}gico (CNPq), the Coordena\c c\~ao de Aperfei\c coamento de Pessoal de N\'{\i}vel Superior (CAPES), and the Funda\c c\~{a}o de Amparo \`{a} Pesquisa do Estado do Rio de Janeiro (FAPERJ) are acknowledged for their financial support.

\end{document}